\documentstyle[aps,prl,epsfig,twocolumn]{revtex}

\draft

\begin{document}
\title{Long-distance quantum communication with atomic ensembles and linear optics}
\author{L.-M. Duan$^{1,2}$\thanks{%
Email: luming.duan@uibk.ac.at}, M. Lukin$^3$, J. I. Cirac$^1$\thanks{%
Email: Ignacio.Cirac@uibk.ac.at}, and P. Zoller$%
^1$}
\address{$^{1}$Institut f\"{u}r Theoretische Physik, Universit\"{a}t Innsbruck,
A-6020 Innsbruck, Austria \\
$^{2}$Laboratory of Quantum Communication and Computation,
USTC, Hefei 230026, China\\
$^{3}$ITAMP, Harvard-Smithsonian Center for Astrophysics,
Cambridge, MA~~02138}
\maketitle

\begin{abstract}
{\bf Quantum communication holds a promise for absolutely secure
transmission of secret messages and faithful transfer of unknown quantum
states. Photonic channels appear to be very attractive for physical
implementation of quantum communication. However, due to losses and
decoherence in the channel, the communication fidelity decreases
exponentially with the channel length. We describe a scheme that allows to
implement robust quantum communication over long lossy channels. The scheme
involves laser manipulation of atomic ensembles, beam splitters, and
single-photon detectors with moderate efficiencies, and therefore well fits
the status of the current experimental technology. We show that the
communication efficiency scale polynomially with the channel length thereby
facilitating scalability to very long distances. }
\end{abstract}

The goal of quantum communication is to transmit quantum states between
distant sites. One the one hand, this has an important potential application
for secret transfer of classical messages by means of quantum cryptography
\cite{1}. On the other hand, it is also an essential element required for
constructing quantum networks. The basic problem of quantum communication is
to generate nearly perfect entangled states between distant sites. Such
states can be used, for example, to implement secure quantum cryptography
using the Ekert protocol \cite{1}, and to faithfully transfer quantum states
via quantum teleportation \cite{2}. All realistic schemes for quantum
communication are presently based on the use of the photonic channels.
However, the degree of entanglement generated between two distant sites
normally decreases exponentially with the length of the connecting channel
due to the optical absorption and other channel noise. To regain a high
degree of entanglement purification schemes can be used \cite{3}. However,
entanglement purification does not fully solve the long-distance
communication problem. Due to the exponential decay of the entanglement in
the channel, one needs an exponentially large number of partially entangled
states to obtain one highly entangled state, which means that for a
sufficiently long distance the task becomes nearly impossible.

To overcome the difficulty associated with the exponential fidelity decay,
the concept of quantum repeaters can be used \cite{4}. In principle, it
allows to make the overall communication fidelity very close to the unity,
with the communication time growing only polynomially with the transmission
distance. In analogy to a fault-tolerant quantum computing \cite{5,6} the
quantum repeater proposal is a cascaded entanglement purification protocol
for communication systems. The basic idea is to divide the transmission
channel into many segments, with the length of each segment comparable to
the channel attenuation length. First, one generates entanglement and
purifies it for each segment; the purified entanglement is then extended to
a longer length by connecting two adjacent segments through entanglement
swapping \cite{2,7}. After entanglement swapping, the overall entanglement
is decreased, and one has to purify it again. One can continue the rounds of
the entanglement swapping and purification until a nearly perfect entangled
states are created between two distance sites.

To implement the quantum repeater protocol, one needs to generate
entanglement between distant quantum bits (qubits), store them for
sufficiently long time and perform local collective operations on several of
these qubits. The requirement of quantum memory is essential since all
purification protocols are probabilistic. When entanglement purification is
performed for each segment of the channel, quantum memory can be used to
keep the segment state if the purification succeeds and to repeat the
purification for the segments only where the previous attempt fails. This is
essentially important for polynomial scaling properties of the communication
efficiency since with no available memory we have to require that the
purifications for all the segments succeeds at the same time; the
probability of such event decreases exponentially with the channel length.
The requirement of quantum memory implies that we need to store the local
qubits in the atomic internal states instead of the photonic states since it
is difficult to store photons for a reasonably long time. With atoms as the
local information carriers it seems to be very hard to implement quantum
repeaters since normally one needs to achieve the strong coupling between
atoms and photons with high-finesse cavities for atomic entanglement
generation, purification, and swapping \cite{8,9}, which, in spite of the
recent significant experimental advances \cite{10,11,12}, remains a very
challenging technology.

Here, we propose a very different scheme which realizes quantum repeaters
and long-distance quantum communication with surprisingly simple physical
setups. The scheme is a combination of three significant advances for
entanglement generation, connection, and applications, with each of the
steps having built-in entanglement purification and resilient to the
realistic noise. The scheme for the fault-tolerant entanglement generation
originates from the earlier proposals to entangle single atoms through
single-photon interference at photodetectors \cite{13,13p}. However, the
present approach involves collective rather than single particle excitations
in atomic ensembles, which allows to significantly simply the realization
and greatly improve the generation efficiency. This is the case due to
collectively enhanced coupling to light that has been recently investigated
both theoretically \cite{14,15,16,17,18} and experimentally \cite{19,20}.
The entanglement connection is achieved through simple linear optical
operations, and is inherently robust against the realistic imperfections.
Different schemes with linear optics are proposed recently for quantum
computation \cite{21} and purification \cite{22}. Finally, the resulting
state of ensembles after the entanglement connection finds direct
applications in realizing the entanglement-based quantum communication
protocols, such as quantum teleportation, cryptography, and Bell inequality
detection. In all of these applications the mixed entanglement is purified
automatically to the nearly perfect entanglement. As a combination of these
three breakthroughs, our scheme circumvents the realistic noise and
imperfections and provides a feasible method for long-distance high-fidelity
quantum communication. The required overhead in the communication time
increases with the distance only polynomially.

\bigskip {\bf Entanglement generation}

The basic element of our system is a cloud of $N_{a}$ identical atoms with
the relevant level structure shown in Fig. 1. A pair of metastable lower
states $|g\rangle $ and $|s\rangle $ can correspond e.g. to hyperfine or
Zeeman sublevels of electronic ground state of alkali atoms. Long lifetimes
for relevant coherence have been observed both in a room-temperature dilute
atomic gas (e.g. in \cite{18,19}) and in a sample of cold trapped atoms
(e.g. in \cite{20}). To facilitate enhanced coupling to light, the atomic
medium is preferably optically thick along one direction. This can be easily
achieved either by working with a pencil shaped atomic sample \cite{18,19,20}
or by placing the sample in a low-finesse ring cavity \cite{15,QND} (see
Supplementary information).

All the atoms are initially prepared in the ground state $\left|
g\right\rangle $. A sample is illuminated by a short, off-resonant laser
pulse that induces Raman transitions into the states $\left| s\right\rangle $%
. We are particularly interested in the forward-scattered Stokes light that
is co-propagating with the laser. Such scattering events are uniquely
correlated with the excitation of the symmetric collective atomic mode $S$
\cite{14,15,16,17,18,19,20} given by $S\equiv \left( 1/\sqrt{N_{a}}\right)
\sum_{i}\left| g\right\rangle _{i}\left\langle s\right| $, where the
summation is taken over all the atoms. In particular, an emission of the
single Stokes photon in a forward direction results in the state of atomic
ensemble given by $S^{\dagger }|0_{a}\rangle $, where the ensemble ground
state $\left| 0_{a}\right\rangle \equiv \bigotimes_{i}\left| g\right\rangle
_{i}$).

We assume that the light-atom interaction time $t_{\Delta }$ is short so
that the mean photon number in the forward-scattered Stokes pulse is much
smaller than $1$. One can define an effective single-mode bosonic operator $%
a $ for this Stokes pulse with the corresponding vacuum state denoted by $%
\left| 0_{p}\right\rangle $. The whole state of the atomic collective mode
and the forward-scattering Stokes mode can now be written in the following
form (see the Supplementary information for the technical details)

\begin{equation}
\left| \phi \right\rangle =\left| 0_{a}\right\rangle \left|
0_{p}\right\rangle +\sqrt{p_{c}}S^{\dagger }a^{\dagger }\left|
0_{a}\right\rangle \left| 0_{p}\right\rangle +o\left( p_{c}\right) ,
\end{equation}
where $p_{c}$ is the small excitation probability and $o\left( p_{c}\right) $%
represents the terms with more excitations whose probabilities are equal or
smaller than $p_{c}^{2}$. Before proceeding we note that a fraction of light
is emitted in other directions due to the spontaneous emissions. However
whenever $N_{a}$ is large, the contribution to the population in the
symmetric collective mode from the spontaneous emissions is small \cite
{14,15,16,17,18,19,20}. As a result we have a large signal-to-noise ratio
for the processes involving the collective mode, which greatly enhances the
efficiency of the present scheme (see Box 1 and the Supplementary
information).

We now show how to use this setup to generate entanglement between two
distant ensembles L and R using the configuration shown in Fig. 1. Here two
laser pulses excited both ensembles simultaneously and the whole system is
described by the state $\left| \phi \right\rangle _{L}\otimes \left| \phi
\right\rangle _{R}$, where $\left| \phi \right\rangle _{L}$ and $\left| \phi
\right\rangle _{R}$ are given by Eq. (1) with all the operators and states
distinguished by the subscript L or R. The forward scattered Stokes light
from both ensembles is combined at the beam splitter and a photodetector
click in either D1 {\it or} D2 measures the combined radiation from two
samples, $a_{+}^{\dagger }a_{+}$ or $a_{-}^{\dagger }a_{-}$ with $a_{\pm
}=\left( a_{L}\pm e^{i\varphi }a_{R}\right) /\sqrt{2}$. Here, $\varphi $
denotes an unknown difference of the phase shifts in the two-side channels.
We can also assume that $\varphi $ has an imaginary part to account for the
possible asymmetry of the setup, which will also be corrected automatically
in our scheme. But the setup asymmetry can be easily made very small, and
for simplicity of expressions we assume $\varphi $ is real in the following.
Conditional on the detector click, we should apply $a_{+}$ or $a_{-}$ to the
whole state $\left| \phi \right\rangle _{L}\otimes \left| \phi \right\rangle
_{R}$, and the projected state of the ensembles L and R is nearly maximally
entangled with the form (neglecting the high-order terms $o\left(
p_{c}\right) $)
\begin{equation}
\left| \Psi _{\varphi }\right\rangle _{LR}^{\pm }=\left( S_{L}^{\dagger }\pm
e^{i\varphi }S_{R}^{\dagger }\right) /\sqrt{2}\left| 0_{a}\right\rangle
_{L}\left| 0_{a}\right\rangle _{R}.
\end{equation}
The probability for getting a click is given by $p_{c}$ for each round, so
we need repeat the process about $1/p_{c}$ times for a successful
entanglement preparation, and the average preparation time is given by $%
T_{0}\sim t_{\Delta }/p_{c}$. The states $\left| \Psi _{r}\right\rangle
_{LR}^{+}$ and $\left| \Psi _{r}\right\rangle _{LR}^{-}$ can be easily
transformed to each other by a simple local phase shift. Without loss of
generality, we assume in the following that we generate the entangled state $%
\left| \Psi _{r}\right\rangle _{LR}^{+}$.

As will be shown below, the presence of the noise modifies the projected
state of the ensembles to
\begin{equation}
\rho _{LR}\left( c_{0},\varphi \right) =\frac{1}{c_{0}+1}\left( c_{0}\left|
0_{a}0_{a}\right\rangle _{LR}\left\langle 0_{a}0_{a}\right| +\left| \Psi
_{\varphi }\right\rangle _{LR}^{\text{ }+}\left\langle \Psi _{\varphi
}\right| \right) ,
\end{equation}
where the ``vacuum'' coefficient $c_{0}$ is determined by the dark count
rates of the photon detectors. It will be seen below that any state in the
form of Eq. (3) will be purified automatically to a maximally entangled
state in the entanglement-based communication schemes. We therefore call
this state an effective maximally entangled (EME) state with the vacuum
coefficient $c_{0}$ determining the purification efficiency.

\bigskip {\bf Entanglement connection through swapping}
%\section{Entanglement connection through swapping}

After the successful generation of the entanglement within the attenuation
length, we want to extend the quantum communication distance. This is done
through entanglement swapping with the configuration shown in Fig. 2.
Suppose that we start with two pairs of the entangled ensembles described by
the state $\rho _{LI_{1}}\otimes \rho _{I_{2}R}$, where $\rho _{LI_{1}}$ and
$\rho _{I_{2}R}$ are given by Eq. (3). In the ideal case, the setup shown in
Fig. 2 measures the quantities corresponding to operators $S_{\pm }^{\dagger
}S_{\pm }$ with $S_{\pm }=\left( S_{I_{1}}\pm S_{I_{2}}\right) /\sqrt{2}$.
If the measurement is successful (i.e., one of the detectors registers one
photon), we will prepare the ensembles L and R into another EME state. The
new $\varphi $-parameter is given by $\varphi _{1}+\varphi _{2}$, where $%
\varphi _{1}$ and $\varphi _{2}$ denote the old $\varphi $-parameters for
the two segment EME states. As will be seen below, even in the presence of
the realistic noise and imperfections, an EME state is still created after a
detector click. The noise only influences the success probability to get a
click and the new vacuum coefficient in the EME state. In general we can
express the success probability $p_{1}$ and the new vacuum coefficient $%
c_{1} $ as $p_{1}=f_{1}\left( c_{0}\right) $ and $c_{1}=f_{2}\left(
c_{0}\right) $, where the functions $f_{1}$ and $f_{2}$ depend on the
particular noise properties.

The above method for connecting entanglement can be cascaded to arbitrarily
extend the communication distance. For the $i$th ($i=1,2,\cdots ,n$)
entanglement connection, we first prepare in parallel two pairs of ensembles
in the EME states with the same vacuum coefficient $c_{i-1}$ and the same
communication length $L_{i-1}$, and then perform the entanglement swapping
as shown in Fig. 2, which now succeeds with a probability $p_{i}=f_{1}\left(
c_{i-1}\right) $. After a successful detector click, the communication
length is extended to $L_{i}=2L_{i-1}$, and the vacuum coefficient in the
connected EME\ state becomes $c_{i}=f_{2}\left( c_{i-1}\right) $. Since the $%
i$th entanglement connection need be repeated in average $1/p_{i}$ times,
the total time needed to establish an EME state over the distance $%
L_{n}=2^{n}L_{0}$ is given by $T_{n}=T_{0}\prod_{i=1}^{n}\left(
1/p_{i}\right) $, where $L_{0}$ denotes the distance of each segment in the
entanglement generation.

\bigskip {\bf Entanglement-based communication schemes}
%\section{Entanglement-based communication schemes}

After an EME\ state has been established between two distant sites, we would
like to use it in the communication protocols, such as quantum
teleportation, cryptography, and Bell inequality detection. It is not
obvious that the EME state (3), which is entangled in the Fock basis, is
useful for these tasks since in the Fock basis it is experimentally hard to
do certain single-bit operations. In the following we will show how the EME\
states can be used to realize all these protocols with simple experimental
configurations.

Quantum cryptography and the Bell inequality detection are achieved with the
setup shown by Fig. 3a. The state of the two pairs of ensembles is expressed
as $\rho _{L_{1}R_{1}}\otimes \rho _{L_{2}R_{2}}$, where $\rho _{L_{i}R_{i}}$
$\left( i=1,2\right) $ denote the same EME state with the vacuum coefficient
$c_{n}$ if we have done $n$ times entanglement connection. The $\varphi $%
-parameters in $\rho _{L_{1}R_{1}}$ and $\rho _{L_{2}R_{2}}$ are the same
provided that the two states are established over the same stationary
channels. We register only the coincidences of the two-side detectors, so
the protocol is successful only if there is a click on each side. Under this
condition, the vacuum components in the EME states, together with the state
components $S_{L_{1}}^{\dagger }S_{L_{2}}^{\dagger }\left| \text{vac}%
\right\rangle $ and $S_{R_{1}}^{\dagger }S_{R_{2}}^{\dagger }\left| \text{vac%
}\right\rangle $, where $\left| \text{vac}\right\rangle $ denotes the
ensemble state $\left| 0_{a}0_{a}0_{a}0_{a}\right\rangle
_{L_{1}R_{1}L_{2}R_{2}}$, have no contributions to the experimental results.
So, for the measurement scheme shown by Fig. 3, the ensemble state $\rho
_{L_{1}R_{1}}\otimes \rho _{L_{2}R_{2}}$ is effectively equivalent to the
following ``polarization'' maximally entangled (PME) state (the terminology
of ``polarization'' comes from an analogy to the optical case)
\begin{equation}
\left| \Psi \right\rangle _{\text{PME}}=\left( S_{L_{1}}^{\dagger
}S_{R_{2}}^{\dagger }+S_{L_{2}}^{\dagger }S_{R_{1}}^{\dagger }\right) /\sqrt{%
2}\left| \text{vac}\right\rangle .
\end{equation}
The success probability for the projection from $\rho _{L_{1}R_{1}}\otimes
\rho _{L_{2}R_{2}}$ to $\left| \Psi \right\rangle _{\text{PME}}$ (i.e., the
probability to get a click on each side) is given by $p_{a}=1/[2\left(
c_{n}+1\right) ^{2}]$. One can also check that in Fig. 3, the phase shift $%
\psi _{\Lambda }$ $\left( \Lambda =L\text{ or }R\right) $ together with the
corresponding beam splitter operation are equivalent to a single-bit
rotation in the basis $\left\{ \left| 0\right\rangle _{\Lambda }\equiv
S_{\Lambda _{1}}^{\dagger }\left| 0_{a}0_{a}\right\rangle _{\Lambda
_{1}\Lambda _{2}},\text{ }\left| 1\right\rangle _{\Lambda }\equiv S_{\Lambda
_{2}}^{\dagger }\left| 0_{a}0_{a}\right\rangle _{\Lambda _{1}\Lambda
_{2}}\right\} $ with the rotation angle $\theta =\psi _{\Lambda }/2$. Now,
it is clear how to do quantum cryptography and Bell inequality detection
since we have the PME state and we can perform the desired single-bit
rotations in the corresponding basis. For instance, to distribute a quantum
key between the two remote sides, we simply choose $\psi _{\Lambda }$
randomly from the set $\left\{ 0,\pi /2\right\} $ with an equal probability,
and keep the measurement results (to be $0$ if $D_{1}^{\Lambda }$ clicks,
and $1$ if $D_{1}^{\Lambda }$ clicks) on both sides as the shared secret key
if the two sides become aware that they have chosen the same phase shift
after the public declare. This is exactly the Ekert scheme \cite{1} and its
absolute security follows directly from the proofs in \cite{23,24}. For the
Bell inequality detection, we infer the correlations $E\left( \psi _{L},\psi
_{R}\right) \equiv
P_{D_{1}^{L}D_{1}^{R}}+P_{D_{2}^{L}D_{2}^{R}}-P_{D_{1}^{L}D_{2}^{R}}-P_{D_{2}^{L}D_{1}^{R}}=\cos \left( \psi _{L}-\psi _{R}\right)
$ from the measurement of the coincidences $P_{D_{1}^{L}D_{1}^{R}}$ etc. For
the setup shown in Fig. 3a, we would have $\left| E\left( 0,\pi /4\right)
+E\left( \pi /2,\pi /4\right) +E\left( \pi /2,3\pi /4\right) -E\left( 0,3\pi
/4\right) \right| =2\sqrt{2}$, whereas for any local hidden variable
theories, the CHSH inequality \cite{25} implies that this value should be
below $2$.

We can also use the established long-distance EME states for faithful
transfer of unknown quantum states through quantum teleportation, with the
setup shown by Fig. 3b. In this setup, if two detectors click on the left
side, there is a significant probability that there is no collective
excitation on the right side since the product of the EME states $\rho
_{L_{1}R_{1}}\otimes \rho _{L_{2}R_{2}}$ contains vacuum components.
However, if there is a collective excitation appearing from the right side,
its ``polarization'' state would be exactly the same as the one input from
the left. So, as in the Innsbruck experiment \cite{26}, the teleportation
here is probabilistic and needs posterior confirmation; but if it succeeds,
the teleportation fidelity would be nearly perfect since in this case the
entanglement is equivalently described by the PME state (4). The success
probability for the teleportation is also given by $p_{a}=1/[2\left(
c_{n}+1\right) ^{2}]$, which determines the average number of repetitions
for a successful teleportation.

\bigskip {\bf Noise and built-in entanglement purification}
%\section{Noise and built-in entanglement purification}

We next discuss noise and imperfections in our schemes for entanglement
generation, connection, and applications. In particular we show that each
step contains built-in entanglement purification which makes the whole
scheme resilient to the realistic noise and imperfections.

In the entanglement generation, the dominant noise is the photon loss, which
includes the contributions from the channel attenuation, the spontaneous
emissions in the atomic ensembles (which results in the population of the
collective atomic mode with the accompanying photon going to other
directions), the coupling inefficiency of the Stokes light into and out of
the channel, and the inefficiency of the single-photon detectors. The loss
probability is denoted by $1-\eta _{p}$ with the overall efficiency $\eta
_{p}=\eta _{p}^{\prime }e^{-L_{0}/L_{\text{att}}}$, where we have separated
the channel attenuation $e^{-L_{0}/L_{\text{att}}}$ ($L_{\text{att}}$ is the
channel attenuation length) from other noise contributions $\eta
_{p}^{\prime }$ with $\eta _{p}^{\prime }$ independent of the communication
distance $L_{0}$. The photon loss decreases the success probably for getting
a detector click from $p_{c}$ to $\eta _{p}p_{c}$, but it has no influence
on the resulting EME state. Due to this noise, the entanglement preparation
time should be replaced by $T_{0}\sim t_{\Delta }/\left( \eta
_{p}p_{c}\right) $. The second source of noise comes from the dark counts of
the single-photon detectors. The dark count gives a detector click, but
without population of the collective atomic mode, so it contributes to the
vacuum coefficient in the EME state. If the dark count comes up with a
probability $p_{dc}$ for the time interval $t_{\Delta }$, the vacuum
coefficient is given by $c_{0}=p_{dc}/\left( \eta _{p}p_{c}\right) $, which
is typically much smaller than $1$ since the Raman transition rate is much
larger than the dark count rate. The final source of noise, which influences
the fidelity to get the EME state, is caused by the event that more than one
atom are excited to the collective mode $S$ whereas there is only one click
in D1 or D2. The conditional probability for that event is given by $p_{c}$,
so we can estimate the fidelity imperfection $\Delta F_{0}\equiv 1-F_{0}$
for the entanglement generation by
\begin{equation}
\Delta F_{0}\sim p_{c}.
\end{equation}
Note that by decreasing the excitation probability $p_{c}$, one can make the
fidelity imperfection closer and closer to zero with the price of a longer
entanglement preparation time $T_{0}$. This is the basic idea of the
entanglement purification. So, in this scheme, the confirmation of the click
from the single-photon detector generates and purifies entanglement at the
same time.

In the entanglement swapping, the dominant noise is still the losses, which
include the contributions from the detector inefficiency, the inefficiency
of the excitation transfer from the collective atomic mode to the optical
mode \cite{19,20}, and the small decay of the atomic excitation during the
storage \cite{18,19,20}. Note that by introducing the detector inefficiency,
we have automatically taken into account the imperfection that the detectors
cannot distinguish the single and the two photons. With all these losses,
the overall efficiency in the entanglement swapping is denoted by $\eta _{s}$%
. The loss in the entanglement swapping gives contributions to the vacuum
coefficient in the connected EME state, since in the presence of loss a
single detector click might result from two collective excitations in the
ensembles I$_{1}$ and I$_{2}$, and in this case, the collective modes in the
ensembles L and R have to be in a vacuum state. After taking into account
the realistic noise, we can specify the success probability and the new
vacuum coefficient for the $i$th entanglement connection by the recursion
relations $p_{i}\equiv f_{1}\left( c_{i-1}\right) =\eta _{s}\left( 1-\frac{%
\eta _{s}}{2\left( c_{i-1}+1\right) }\right) /\left( c_{i-1}+1\right) $ and $%
c_{i}\equiv f_{2}\left( c_{i-1}\right) =2c_{i-1}+1-\eta _{s}$. The
coefficient $c_{0}$ for the entanglement preparation is typically much
smaller than $1-\eta _{s}$, then we have $c_{i}\approx \left( 2^{i}-1\right)
\left( 1-\eta _{s}\right) =(L_{i}/L_{0}-1)\left( 1-\eta _{s}\right) $, where
$L_{i}$ denotes the communication distance after $i$ times entanglement
connection. With the expression for the $c_{i}$, we can easily evaluate the
probability $p_{i}$ and the communication time $T_{n}$ for establishing a
EME state over the distance $L_{n}=2^{n}L_{0}$. After the entanglement
connection, the fidelity of the EME state also decreases, and after $n$
times connection, the overall fidelity imperfection $\Delta F_{n}\sim
2^{n}\Delta F_{0}\sim \left( L_{n}/L_{0}\right) \Delta F_{0}$. We need fix $%
\Delta F_{n}$ to be small by decreasing the excitation probability $p_{c}$
in Eq. (5).

It is important to point out that our entanglement connection scheme also
has built-in entanglement purification function. This can be understood as
follows: Each time we connect entanglement, the imperfections of the setup
decrease the entanglement fraction $1/\left( c_{i}+1\right) $ in the EME
state. However, the entanglement fraction decays only linearly with the
distance (the number of segments), which is in contrast to the exponential
decay of the entanglement for the connection schemes without entanglement
purification. The reason for the slow decay is that in each time of the
entanglement connection, we need repeat the protocol until there is a
detector click, and the confirmation of \ a click removes part of the added
vacuum noise since a larger vacuum components in the EME state results in
more times of repetitions. The built-in entanglement purification in the
connection scheme is essential for the polynomial scaling law of the
communication efficiency.

As in the entanglement generation and connection schemes, our entanglement
application schemes also have built-in entanglement purification which makes
them resilient to the realistic noise. Firstly, we have seen that the vacuum
components in the EME states are removed from the confirmation of the
detector clicks and thus have no influence on the fidelity of all the
application schemes. Secondly, if the single-photon detectors and the
atom-to-light excitation transitions in the application schemes are
imperfect with the overall efficiency denoted by $\eta _{a}$, one can easily
check that these imperfections only influence the efficiency to get the
detector clicks with the success probability replaced by $p_{a}=\eta _{a}/%
\left[ 2\left( c_{n}+1\right) ^{2}\right] $, and have no effects on the
communication fidelity. Finally, we have seen that the phase shifts in the
stationary channels and the small asymmetry of the stationary setup are
removed automatically when we project the EME state to the PME state, and
thus have no influence on the communication fidelity.

The noise not correctable by our scheme includes the detector dark count in
the entanglement connection and the non-stationary channel noise and set
asymmetries. The resulting fidelity imperfection from the dark count
increases linearly with the number of segments $L_{n}/L_{0}$, and form the
non-stationary channel noise and set asymmetries increases by the
random-walk law $\sqrt{L_{n}/L_{0}}$. For each time of entanglement
connection, the dark count probability is about $10^{-5}$ if we make a
typical choice that the collective emission rate is about $10$MHz and the
dark count rate is $10^{2}$Hz. So this noise is negligible even if we have
communicated over a long distance ($10^{3}$ the channel attenuation length $%
L_{\text{att}}$ for instance). The non-stationary channel noise and setup
asymmetries can also be safely neglected for such a distance. For instance,
it is relatively easy to control the non-stationary asymmetries in local
laser operations to values below $10^{-4}$ with the use of accurate
polarization techniques \cite{28} for Zeeman sublevels (as in Fig. 2b).

\bigskip {\bf Scaling of the communication efficiency}
%\section{Scaling of the communication efficiency}

We have shown that each of our entanglement generation, connection, and
application schemes has built-in entanglement purification, and as a result
of this property, we can fix the communication fidelity to be nearly
perfect, and at the same time keep the communication time to increase only
polynomially with the distance. Assume that we want to communicate over a
distance $L=L_{n}=2^{n}L_{0}$. By fixing the overall fidelity imperfection
to be a desired small value $\Delta F_{n}$, the entanglement preparation
time becomes $T_{0}\sim t_{\Delta }/\left( \eta _{p}\Delta F_{0}\right) \sim
\left( L_{n}/L_{0}\right) t_{\Delta }/\left( \eta _{p}\Delta F_{n}\right) $.
For an effective generation of the PME state (4), the total communication
time $T_{\text{tot}}\sim T_{n}/p_{a}$ $\ $with $T_{n}\sim
T_{0}\prod_{i=1}^{n}\left( 1/p_{i}\right) $. So the total communication time
scales with the distance by the law
\begin{equation}
T_{\text{tot}}\sim 2\left( L/L_{0}\right) ^{2}/\left( \eta _{p}p_{a}\Delta
F_{T}\Pi _{i=1}^{n}p_{i}\right) ,
\end{equation}
where the success probabilities $p_{i},p_{a}$ for the $i$th entanglement
connection and for the entanglement application have been specified before.
The expression (6) has confirmed that the communication time $T_{\text{tot}}$
increases with the distance $L$ only polynomially. We show this explicitly
by taking two limiting cases. In the first case, the inefficiency $1-\eta
_{s}$ for the entanglement swapping is assumed to be negligibly small. One
can deduce from Eq. (6) that in this case the communication time $T_{\text{%
tot}}\sim T_{\text{con}}\left( L/L_{0}\right) ^{2}e^{L_{0}/L_{\text{att}}}$,
with the constant $T_{\text{con}}\equiv 2t_{\Delta }/\left( \eta
_{p}^{\prime }\eta _{a}\Delta F_{T}\right) $ being independent of the
segment and the total distances $L_{0}$ and $L$. The communication time $T_{%
\text{tot}}$ increases with $L$ quadratically. In the second case, we assume
that the inefficiency $1-\eta _{s}$ is considerably large. The communication
time in this case is approximated by $T_{\text{tot}}\sim T_{\text{con}%
}(L/L_{0})^{[\log _{2}\left( L/L_{0}\right) +1]/2+\log _{2}(1/\eta
_{s}-1)+2}e^{L_{0}/L_{\text{att}}}$, which increases with $L$ still
polynomially (or sub-exponentially in a more accurate language, but this
makes no difference in practice since the factor $\log _{2}\left(
L/L_{0}\right) $ is well bounded from above for any reasonably long
distance). If $T_{\text{tot}}$ increases with $L/L_{0}$ by the $m$th power
law $\left( L/L_{0}\right) ^{m}$, there is an optimal choice of the segment
length to be $L_{0}=mL_{\text{att}}$ to minimize the time $T_{\text{tot}}$.
As a simple estimation of the improvement in the communication efficiency,
we assume that the total distance $L$ is about $100L_{\text{att}}$, for a
choice of the parameter $\eta _{s}\approx 2/3$, the communication time $T_{%
\text{tot}}/T_{\text{con}}\sim 10^{6}$ with the optimal segment length $%
L_{0}\sim 5.7L_{\text{att}}$. This result is a dramatic improvement compared
with the direct communication case, where the communication time $T_{\text{%
tot}}$ for getting a PME state increases with the distance $L$ by the
exponential law $T_{\text{tot}}\sim T_{\text{con}}e^{L/L_{\text{att}}}$. For
the same distance $L\sim 100L_{\text{att}}$, one needs $T_{\text{tot}}/T_{%
\text{con}}\sim 10^{43}$ for direct communication, which means that for this
example the present scheme is $10^{37}$ times more efficient .

\bigskip {\bf Outlook} %\section{Outlook}

We have presented a novel approach for implementation of quantum repeaters
and long-distance quantum communication. The proposed technique allows to
generate and connect the entanglement and use it in quantum teleportation,
cryptography, and tests of Bell inequalities. All of the elements of the
present scheme are within the reach of current experimental technology, and
all have the important property of built-in entanglement purification which
makes them resilient to the realistic noise. As a result, the overhead
required to implement the present scheme, such as the communication time,
scales polynomially with the channel length. This is in dramatic contrast to
direct communication where the exponential overhead is required. Such an
efficient scaling, combined with a relative simplicity of the experimental
setup, opens up realistic prospective for quantum communication over long
distances.

{\bf Acknowledgments} This work was supported by the Austrian Science
Foundation, the Europe Union project EQUIP, the ESF, the European TMR
network Quantum Information, and the NSF through the grant to the ITAMP.
L.M.D. thanks in addition the support from the Chinese Science Foundation.

\newpage

{\bf Box 1: Collective enhancement}

Long-lived excitations in atomic ensemble can be viewed as waves of excited
spins. We are here particularly interested in symmetric spin wave mode $S$.
For a simple demonstration of the collective enhancement, we assume that
atoms are placed in a low finesse ring cavity \cite{QND}, with a relevant
cavity mode corresponding to forward-scattered Stokes radiation. Cavity-free
case corresponds to the limit where the finesse tends to $1$ \cite{15}. The
interaction between the forward-scattered light mode and atoms is described
by the Hamiltonian

\[
H=\hbar \left( \sqrt{N_{a}}\Omega g_{c}/\Delta \right) S^{\dagger
}b^{\dagger }+\text{h.c.},
\]
where $b^{\dagger }$ is creation operator for cavity photon, $\Omega $ is
the laser Rabi frequency, and $g_{c}$ atom-field coupling constant. In
addition to coherent evolution the photonic field mode can leak out of the
cavity at a rate $\kappa $, whereas atomic coherence is dephased by
spontaneous photon scattering into random directions that occurs at a rate $%
\gamma _{s}^{\prime }=\Omega ^{2}/\Delta ^{2}\gamma _{s}$ for each atom,
with $\gamma _{s}$ being the natural linewidth of the electronic excited
state. We emphasize that in the absence of superradiant effects spontaneous
emission events are independent for each atom.

In the bad-cavity limit, we can adiabatically eliminate the cavity mode, and
the resulting dynamics for the collective atomic mode is described by the
Heisenberg-Langevin equation (see the supplementary information for details)

\[
\stackrel{.}{S^{\dagger }}=\frac{(\kappa ^{\prime }-\gamma _{s}^{\prime })}{2%
}S^{\dagger }-\sqrt{\kappa ^{\prime }}b_{in}\left( t\right) +{\rm noise},
\]
where $\kappa ^{\prime }=4|\Omega |^{2}g_{c}^{2}N_{a}/(\Delta ^{2}\kappa ),$
$b_{in}$ is a vacuum field leaking into the cavity, and the last term
represents the fluctuating noise field corresponding to the spontaneous
emission. Note that nature of the dynamics is determined by ratio between
the build-up of coherence due to forward-scattered photons $\kappa ^{\prime }
$ and coherence decay due to spontaneous emission $\gamma _{s}^{\prime }$.
The signal-to-noise ratio is therefore given by $R=\kappa
^{\prime }/\gamma _{s}\equiv 4N_{a}g_{c}^{2}/(\kappa \gamma )$, which is
large when many-atom ensemble is used. In the cavity-free case this
expression corresponds to optical depth (density-length product) of the
sample. The result should be compared with the signal-to-noise ratio in the
single-atom case $N_{a}=1$, where to obtain $R>1$ a high-Q microcavity is
required \cite{10,11,12}. The collective enhancement takes place since the
coherent forward scattering involves only one collective atomic mode $S$,
whereas the spontaneous emissions distribute excitation over all atomic
modes. Therefore only a small fraction of spontaneous emission events
influences the symmetric mode $S$, which results in a large signal-to-noise
ratio.

\newpage

Caption for Fig. 1 (1a) The relevant level structure of the atoms in the
ensemble with $\left| g\right\rangle $, the ground state, $\left|
s\right\rangle ,$ the metastable state for storing a qubit, and $\left|
e\right\rangle ,$ the excited state. The transition $\left| g\right\rangle
\rightarrow \left| e\right\rangle $ is coupled by the classical laser with
the Rabi frequency $\Omega $, and the forward scattering Stokes light comes
from the transition $\left| e\right\rangle \rightarrow \left| s\right\rangle
$. For convenience, we assume off-resonant coupling with a large detuning $%
\Delta $. (1b) Schematic setup for generating entanglement between the two
atomic ensembles L and R. The two ensembles are pencil shaped and
illuminated by the synchronized classical laser pulses. The
forward-scattering Stokes pulses are collected after the filters
(polarization and frequency selective) and interfered at a 50\%-50\% beam
splitter BS after the transmission channels, with the outputs detected
respectively by two single-photon detectors D1 and D2. If there is a click
in D1 {\it or} D2, the process is finished and we successfully generate
entanglement between the ensembles L and R. Otherwise, we first apply a
repumping pulse to the transition $\left| s\right\rangle \rightarrow \left|
e\right\rangle $ on the ensembles L and R to set the state of the ensembles
back to the ground state $\left| 0\right\rangle _{a}^{L}\otimes \left|
0\right\rangle _{a}^{R}$, then the same classical laser pulses as the first
round are applied to the transition $\left| g\right\rangle \rightarrow
\left| e\right\rangle $ and we detect again the forward-scattering Stokes
pulses after the beam splitter. This process is repeated until finally we
have a click in the D1 {\it or} D2 detector.

Caption of Fig. 2. (2a) Illustrative setup for the entanglement swapping. We
have two pairs of ensembles L, I$_{1}$ and I$_{2}$, R distributed at three
sites L, I and R. Each of the ensemble-pairs L, I$_{1}$ and I$_{2}$, R is
prepared in an EME state in the form of Eq. (3). The excitations in the
collective modes of the ensembles I$_{1}$ and I$_{2}$ are transferred
simultaneously to the optical excitations by the repumping pulses applied to
the atomic transition $\left| s\right\rangle \rightarrow \left|
e\right\rangle $, and the stimulated optical excitations, after a 50\%-50\%
beam splitter, are detected by the single-photon detectors D1 and D2. If
either D1 {\it or} D2 clicks, the protocol is successful and an EME state in
the form of Eq. (3) is established between the ensembles L and R with a
doubled communication distance. Otherwise, the process fails, and we need to
repeat the previous entanglement generation and swapping until finally we
have a click in D1 or D2, that is, until the protocol finally succeeds. (2b)
The two intermediated ensembles I$_{1}$ and I$_{2}$ can also be replaced by
one ensemble but with two metastable states I$_{1}$ and I$_{2}$ to store the
two different collective modes. The 50\%-50\% beam splitter operation can be
simply realized by a $\pi /2$ pulse on the two metastable states before the
collective atomic excitations are transferred to the optical excitations.

Caption of Fig. 3 (3a) Schematic setup for the realization of quantum
cryptography and Bell inequality detection. Two pairs of ensembles L$_{1}$, R%
$_{1}$ and L$_{2}$, R$_{2}$ (or two pairs of metastable states as shown by
Fig. (2b)) have been prepared in the EME states. The collective atomic
excitations on each side are transferred to the optical excitations, which,
respectively after a relative phase shift $\varphi _{L}$ or $\varphi _{R}$
and a 50\%-50\% beam splitter, are detected by the single-photon detectors $%
D_{1}^{L},D_{2}^{L}$ and $D_{1}^{R},D_{2}^{R}$. We look at the four possible
coincidences of $D_{1}^{R},D_{2}^{R}$ with $D_{1}^{L},D_{2}^{L}$, which are
functions of the phase difference $\varphi _{L}-\varphi _{R}$. Depending on
the choice of $\varphi _{L}$ and $\varphi _{R}$, this setup can realize both
the quantum cryptography and the Bell inequality detection. (3b) Schematic
setup for probabilistic quantum teleportation of the atomic ``polarization''
state. Similarly, two pairs of ensembles L$_{1}$, R$_{1}$ and L$_{2}$, R$%
_{2} $ are prepared in the EME states. We want to teleport an atomic
``polarization'' state $\left( d_{0}S_{I_{1}}^{\dagger
}+d_{1}S_{I_{2}}^{\dagger }\right) \left| 0_{a}0_{a}\right\rangle
_{I_{1}I_{2}}$ with unknown coefficients $d_{0},d_{1}$ from the left to the
right side, where $S_{I_{1}}^{\dagger },S_{I_{2}}^{\dagger }$ denote the
collective atomic operators for the two ensembles I$_{1}$ and I$_{2}$ (or
two metastable states in the same ensemble). The collective atomic
excitations in the ensembles I$_{1}$, L$_{1}$ and I$_{2}$, L$_{2}$ are
transferred to the optical excitations, which, after a 50\%-50\% beam
splitter, are detected by the single-photon detectors $D_{1}^{I},D_{1}^{L}$
and $D_{2}^{I},D_{2}^{L}$. If there are a click in $D_{1}^{I}$ {\it or }$%
D_{1}^{L}$ and a click in $D_{2}^{I}$ {\it or }$D_{2}^{I}$, the protocol is
successful. A $\pi $-phase rotation is then performed on the collective mode
of the ensemble R$_{2}$ conditional on that the two clicks appear in the
detectors $D_{1}^{I}$,$D_{2}^{L}$ or $D_{2}^{I}$,$D_{1}^{L}$. The collective
excitation in the ensembles R$_{1}$ and R$_{2}$, if appearing, would be
found in the same ``polarization'' state $\left( d_{0}S_{R_{1}}^{\dagger
}+d_{1}S_{R_{2}}^{\dagger }\right) \left| 0_{a}0_{a}\right\rangle
_{R_{1}R_{2}}$.

\newpage %\epsfig{file=d61np1.eps,width=8cm}
\begin{figure}[tbp]
\epsfig{file=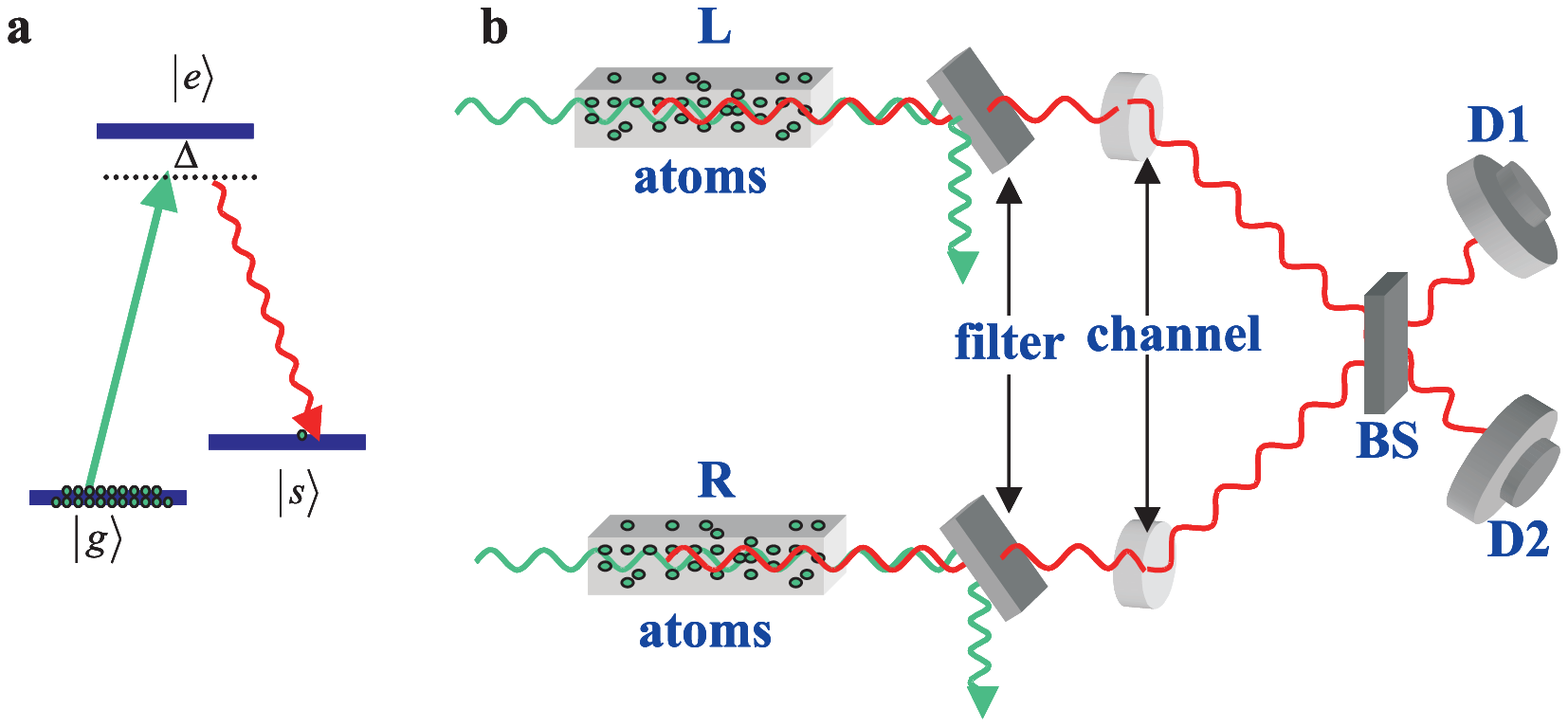,width=8cm} \caption{{}}
\end{figure}

\bigskip

\begin{figure}[tbp]
\epsfig{file=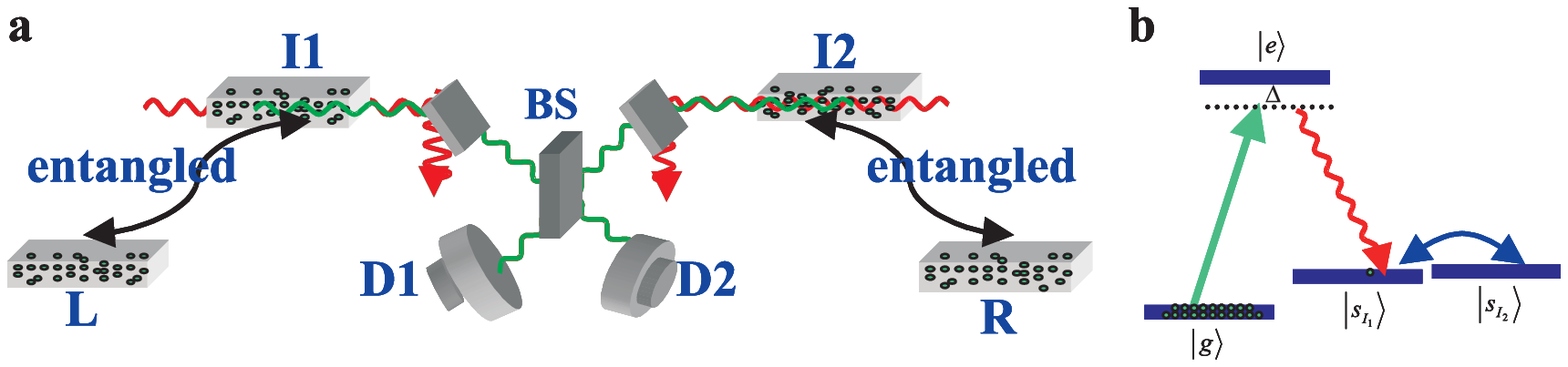,width=8cm} \caption{}
\end{figure}

\bigskip

\begin{figure}[tbp]
\epsfig{file=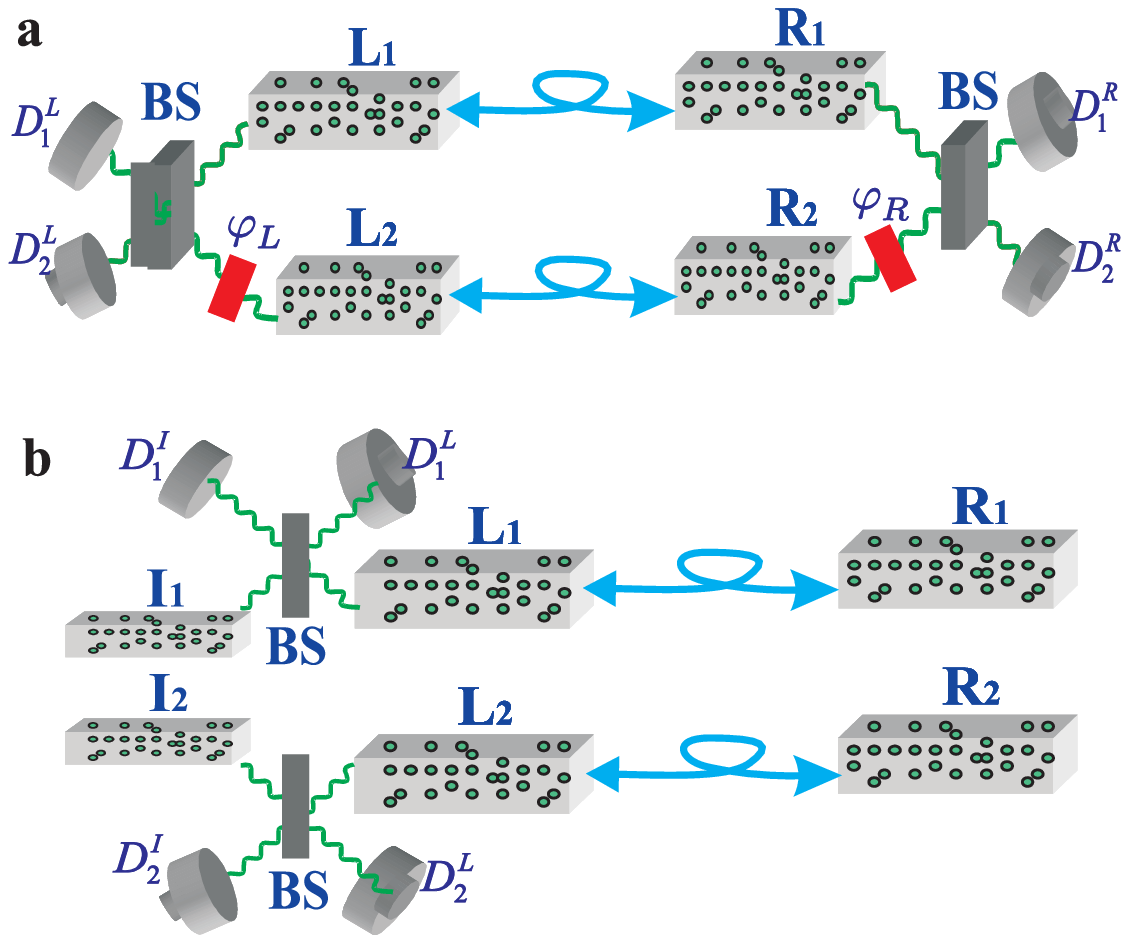,width=6cm} \caption{}
\end{figure}

\end{document}

\title{Long-distance quantum communication with atomic ensembles and linear optics:
\\
Supplementary information\\
}
\author{L.-M. Duan$^{1,2}$\thanks{%
Email: luming.duan@uibk.ac.at}, M. Lukin$^3$, J. I. Cirac$^{1}$\thanks{%
Email: Ignacio.Cirac@uibk.ac.at}, and P. Zoller$%
^1$}
\address{$^{1}$Institut f\"{u}r Theoretische Physik, Universit\"{a}t Innsbruck,
A-6020 Innsbruck, Austria \\
$^{2}$Laboratory of Quantum Communication and Computation,
USTC, Hefei 230026, China\\
$^{3}$ITAMP, Harvard-Smithsonian Center for Astrophysics,
Cambridge, MA~~02138}
\maketitle

\section{Light-atom interaction and the collective enhancement of the
signal-to-noise ratio}

We demonstrate in the following that when optical pulses interact with an
sample of atoms with the $\Lambda $ level configuration, there is a
collective enhancement of the signal-to-noise ratio for the collective
atomic mode. For a simple demonstration of the collective effects, we first
assume there is a (low finesse) ring cavity around the ensemble as shown by
Fig. 1. The case of the free-space ensemble corresponds the limit that the
cavity finesse tends to $1$ [1]. A classical laser with the wave vector $%
k_{l}=\omega _{l}/c$ is coupling to the transition $\left| g\right\rangle
\rightarrow \left| e\right\rangle $ with a Rabi frequency $\Omega $, and the
ring cavity mode $b$ with the wave vector $k_{s}=\omega _{s}/c$ is coupling
to the transition $\left| e\right\rangle \rightarrow \left| s\right\rangle $
and with a coupling coefficient $g_{c}$. For simplicity, first we assume
off-resonant coupling with a large detuning $\Delta $ as shown by Fig. S1.
The classical laser is co-propagating with the ring cavity mode to satisfy
the collective condition $\left( k_{l}-k_{s}\right) L_{a}\ll \pi $, where $%
L_{a}$ is the length of the pencil-shape atomic ensemble. With this
condition, after we adiabatically eliminate the upper level $\left|
e\right\rangle $, the Hamiltonian has the following form in the rotating
frame
\begin{equation}
H=\hbar \left( \sqrt{N_{a}}\Omega g_{c}/\Delta \right) S^{\dagger
}b^{\dagger }+\text{h.c.,}
\end{equation}
where $S$ is the collective atomic mode defined in the text which has the
form $S\equiv \left( 1/\sqrt{N_{a}}\right) \sum_{i}S_{i}$ with $S_{i}=\left|
g\right\rangle _{i}\left\langle s\right| $, the individual atomic lowering
operator. There should be also some light shift terms, but they can be
trivially canceled by refining the laser frequency. Corresponding to the
Hamiltonian (1), the Heisenberg-Langevin equations for the modes $S$ and $b$
are respectively given by [2]
\begin{eqnarray}
\stackrel{.}{b} &=&-i\left( \sqrt{N_{a}}\Omega g_{c}/\Delta \right)
S^{\dagger }-\left( \kappa /2\right) b-\sqrt{\kappa }b_{in}\left( t\right) ,
\\
\stackrel{.}{S^{\dagger }} &=&i\left( \sqrt{N_{a}}\Omega ^{\ast }g_{c}^{\ast
}/\Delta \right) b,
\end{eqnarray}
where $\kappa $ is the cavity decay rate, and $b_{in}\left( t\right) $ is
the input vacuum field with the properties $\left[ b_{in}\left( t\right)
,b_{in}^{\dagger }\left( t^{\prime }\right) \right] =\delta \left(
t-t^{\prime }\right) $ and $\left\langle b_{in}^{\dagger }\left( t\right)
b_{in}\left( t\right) \right\rangle =0$. The output field $b_{out}\left(
t\right) $ from the cavity (the Stokes light) is connected with the input by
the input-output relation $b_{out}\left( t\right) =b_{in}\left( t\right) +%
\sqrt{\kappa }b\left( t\right) $. In obtaining Eq. (3), we have used the
weak interaction condition which assumes after the interaction most of the
atoms are still in the ground state so that $\left[ S,S^{\dagger }\right]
=\sum_{i}\left( \left| g\right\rangle _{i}\left\langle g\right| -\left|
s\right\rangle _{i}\left\langle s\right| \right) /N_{a}\simeq 1$. We have
not included in Eq. (3) the atomic spontaneous emission, which will
introduce some heating effects to the mode $S$ and will be discussed later.
In the bad-cavity limit $\kappa \gg \sqrt{N_{a}}\left| \Omega g_{c}\right|
/\Delta $, we can adiabatically eliminate the cavity mode $b$, and the
resulting equation is
\begin{equation}
\stackrel{.}{S^{\dagger }}=\frac{\kappa ^{\prime }}{2}S^{\dagger }-\sqrt{%
\kappa ^{\prime }}b_{in}\left( t\right) ,
\end{equation}
where the effective interaction coefficient $\kappa ^{\prime }=4N_{a}\left|
\Omega g_{c}\right| ^{2}/\left( \Delta ^{2}\kappa \right) $, and without
loss of generality we assumed the phase of the laser is chosen in the way $%
i\Omega ^{\ast }g_{c}^{\ast }=\left| \Omega g_{c}\right| $. The output
optical field, expressed by the operator $S^{\dagger }$, has the form $%
b_{out}\left( t\right) =-b_{in}\left( t\right) +\sqrt{\kappa ^{\prime }}%
S^{\dagger }\left( t\right) $. Equation (4) is linear and has the explicit
solution
\begin{equation}
S^{\dagger }\left( t\right) =S^{\dagger }\left( 0\right) e^{\kappa ^{\prime
}t/2}-\sqrt{\kappa ^{\prime }}\int_{0}^{t}e^{\kappa ^{\prime }\left( t-\tau
\right) /2}b_{in}\left( \tau \right) d\tau ,
\end{equation}
From this, we easily get the solution for the output optical field $%
b_{out}\left( t\right) $.

What we measure (without beam splitters) by the photon detector is the
integration of the photon current for the detection time interval $t_{\Delta
}$, which is proportional to the intensity integration of the output field.
So we measure the operator $Q_{m}=\int_{0}^{t_{\Delta }}b_{out}^{\dagger
}\left( \tau \right) b_{out}\left( \tau \right) d\tau $. One can find an
explicit expression for $Q_{m}$ by substituting the solution of $%
b_{out}\left( t\right) $. To simplify it, we define an effective single-mode
bosonic operator $a$ from the continuous field $b_{in}\left( t\right) $ with
the form
\begin{equation}
a\equiv -\frac{\sqrt{\kappa ^{\prime }}}{\sqrt{e^{\kappa ^{\prime }t_{\Delta
}}-1}}\int_{0}^{t_{\Delta }}e^{\kappa ^{\prime }\left( t_{\Delta }-\tau
\right) /2}b_{in}\left( \tau \right) d\tau .
\end{equation}
With this operator, the measured quantity is expressed as
\begin{equation}
Q_{m}=a_{t}^{\dagger }a_{t}+\int_{0}^{t_{\Delta }}b_{in}^{\dagger }\left(
\tau \right) b_{in}\left( \tau \right) d\tau ,
\end{equation}
where $a_{t_{\Delta }}=a\cosh r_{c}+S^{\dagger }\left( 0\right) \sinh r_{c},$
a Bogoliubov transformation of $a$ and $S^{\dagger }\left( 0\right) $ with $%
\cosh r_{c}\equiv e^{\kappa ^{\prime }t_{\Delta }/2}$. The last term of Eq.
(7) is a trivial integration of the intensity of the vacuum field, which has
no contribution to the measurement result. So what we measure is in fact the
photon number in the defined effective mode. Note that Eq. (5) can also be
written in the Bogoliubov form $S^{\dagger }\left( t_{\Delta }\right)
=S^{\dagger }\left( 0\right) \cosh r_{c}+a\sinh r_{c}$ with $S^{\dagger
}\left( 0\right) $ and $a$ respectively in the atomic and photonic vacuum
states $\left| 0_{a}\right\rangle $ and $\left| 0_{p}\right\rangle $.
Transferring to the Schrodinger picture, we conclude that after time $%
t_{\Delta }$ the collective atomic mode and the effective mode for the
Stokes light are in a two-mode squeezed state
\begin{equation}
\left| \phi \right\rangle =\sec r_{c}\sum_{n}\left( S^{\dagger }a\tanh
r_{c}\right) ^{n}\left| 0_{a}\right\rangle \left| 0_{p}\right\rangle .
\end{equation}
This is exactly the state (1) in the paper if the excitation probability $%
p_{c}=\tanh ^{2}r_{c}\ll 1$, and we have shown above that the detector
measures the photon number in the effective mode $a$. If there are two
ensembles and the detectors are put after a beam splitter as discussed in
the paper, it is straightforward to extend the above treatment to see that
one measures the photon numbers in the effective modes $a_{\pm }$ (see in
the paper for the definitions). So we confirm the results there.

Now let us take into account the atomic spontaneous emissions from the level
$\left| e\right\rangle $ to $\left| s\right\rangle $. The photons from
spontaneous emissions go to the free-space modes (other than the cavity mode
$b$) with random directions, and we assume that the atomic ensemble is
dilute with $k_{s}/\sqrt[3]{\rho _{n}}\succsim 1$ (where $\rho _{n}$ is the
atomic number density) so that there is no superradiance. Each atom
undergoes spontaneous emissions independently with the rate $\gamma
_{s}^{\prime }=\left( \Omega ^{2}/\Delta ^{2}\right) \gamma _{s},$ where $%
\gamma _{s}$ denotes the on-resonance spontaneous emission rate. The
spontaneous emission introduces a coherence decay term to the Langevin
equation of the individual atomic operator $S_{i}$%
\begin{equation}
\stackrel{.}{S^{\dagger }}_{i}=-\left( \gamma _{s}^{\prime }/2\right)
S_{i}^{\dagger }+\text{noise,}
\end{equation}
where the last term represents the corresponding fluctuation from the noise
field which results in heating, and we have left out in Eq. (9) the coherent
interaction term from the Hamiltonian. By taking summation of Eq. (9) over
all the atoms, we immediately see that there is added coherence decay term
to the Langevin equation (4) of the collective atomic operator $S^{\dagger }$
with the decay rate still given by $\gamma _{s}^{\prime }$. The ratio $R_{sn}
$ between the coherent interaction rate $\kappa ^{\prime }$ and the decay
rate $\gamma _{s}^{\prime }$ (called the signal-to-noise ratio in the
following) is given by $R_{sn}=4N_{a}\left| g_{c}\right| ^{2}/\left( \kappa
\gamma _{s}\right) $. In the single-atom case, the signal-to-noise ratio is
about $4\left| g_{c}\right| ^{2}/\left( \kappa \gamma _{s}\right) $. So, for
the atomic ensemble, the signal-to-noise ratio, which influences the
collection efficiency in our scheme (see the paper for the discussions on
noise), is greatly enhanced by the large factor of the atom number. This
enhancement comes from the fact that the coherent interaction producing the
co-propagating signal involves only the collective atomic mode $S$, whereas
the independent spontaneous emissions distribute over all the atomic modes,
and thus only have small influence on the interesting mode $S$. As a result
of the collective enhancement, a weak-coupling cavity has assured a large
signal-to-noise ratio for the interesting mode.

The collective enhancement of the signal-to-noise ratio can also be easily
understood if one looks at the master equation. The whole density operator $%
\rho _{w}$ for the atomic internal states and the cavity mode obeys the
following master equation [2]
\begin{equation}
\stackrel{.}{\rho }_{w}=i\left[ \rho _{w},H\right] +\kappa \widehat{L}\left[
b\right] \rho _{w}+\gamma _{s}^{\prime }\sum_{i}\widehat{L}\left[
S_{i}^{\dagger }\right] \rho _{w},
\end{equation}
where the Liouville superoperators $\widehat{L}\left[ X\right] $ $\left(
X=b,S_{i}^{\dagger }\right) $ are defined as $\widehat{L}\left[ X\right]
\rho _{w}\equiv X\rho _{w}X^{\dagger }-\left( X^{\dagger }X\rho _{w}+\rho
_{w}X^{\dagger }X\right) /2$. In Eq. (10), the first term of the right hand
side (r.h.s.) comes from the Hamiltonian interaction, the second term
represents the cavity output coupling, and the last term describes
independent spontaneous emissions for individual atomic operators. In the
bad cavity limit, after adiabatically eliminating the cavity mode, we get
from Eqs. (10) and (1) the following master equation for the traced atomic
density operator $\rho _{a}$
\begin{equation}
\stackrel{.}{\rho }_{a}=\kappa ^{\prime }\widehat{L}\left[ S^{\dagger }%
\right] \rho _{a}+\gamma _{s}^{\prime }\sum_{i}\widehat{L}\left[
S_{i}^{\dagger }\right] \rho _{a},
\end{equation}
where $\widehat{L}\left[ S^{\dagger }\right] $ is the Liouville
superoperator for the collective atomic mode. The above equation can be
further simplified if we introduce the Fourier transformation to the
individual atomic operators $S_{j}$ $\left( j=0,1,\cdots ,N_{a}-1\right) $
with the form $S_{\mu }\equiv \sum_{j}S_{j}e^{ij\mu /N_{a}}/\sqrt{N_{a}}$,
where $S_{\mu =0}$ gives exactly the collective atomic operator $S$. In
terms of the operators $S_{\mu }$, the master equation has the form
\begin{equation}
\stackrel{.}{\rho }_{a}=\left( \kappa ^{\prime }+\gamma _{s}^{\prime
}\right) \widehat{L}\left[ S^{\dagger }\right] \rho _{a}+\gamma _{s}^{\prime
}\sum_{\mu \neq 0}\widehat{L}\left[ S_{\mu }^{\dagger }\right] \rho _{a},
\end{equation}
Under the weak interaction condition $\left\langle S_{j}^{\dagger
}S_{j}\right\rangle \ll 1$, the operators $S_{\mu }$ $\left( \mu =0,1,\cdots
,N_{a}-1\right) $ commute with each other, so they represent independent
atomic modes. We are only interested in the collective atomic mode $S$, and
the populations in all the other modes $S_{\mu }$ with $\mu \neq 0$ have no
influence on the state and the measurement of the mode $S$. (To measure the
state of the mode $S$, we transfer the collective atomic excitation to the
optical excitation as described in the paper. The details on the excitation
transferring can be found in Refs. [3,4]). So we can trace over the modes $%
S_{\mu }$ $\left( \mu \neq 0\right) $ and eliminate the last term in Eq.
(12). There are two contributions to the population in the collective atomic
mode: the one with a rate $\kappa ^{\prime }$ produces a coherent output
signal, and the one with a rate $\gamma _{s}^{\prime }$ emits photons to
other random directions. The signal-to-noise ration is again given by $%
R_{sn}=\kappa ^{\prime }/\gamma _{s}^{\prime }\sim 4N_{a}\left| g_{c}\right|
^{2}/\left( \kappa \gamma _{s}\right) $. It is interesting to note from Eq.
(12) that the total spontaneous emission rate of all the modes is $%
N_{a}\gamma _{s}^{\prime }$, which could be much larger than the coherent
interaction rate $\kappa ^{\prime }$, however, the spontaneous emission rate
for the collective atomic mode is $N_{a}$ times smaller than the total rate,
and the spontaneous emissions going to other modes have no influence on the
scheme. So we have a large signal-to-noise ratio. The collective enhancement
of the signal-to-noise ratio has been used in several schemes [1,3-5] and
has been demonstrated by the recent experiments [6-8].

In our scheme by no means we necessarily need a good cavity, since on the
one hand the signal-to-noise ratio in our scheme influences only the
efficiency instead of the fidelity, and on the other hand, due to the
collective enhancement shown above, we could still have a considerably large
signal-to-noise ratio even without a cavity. In fact, we can assume to
continuously decrease the cavity finesse down to $1,$ i.e., to the
free-space limit. In the free-space limit, the cavity decay rate $\kappa $
is estimated by $c/L_{a}$, the inverse of the traveling time of the pulse in
the ensemble. With the well known expressions for the coefficients $g_{c}$
and $\gamma _{s}$ [5], one can estimate the signal-to-noise ratio in the
free-space limit by $R_{sn}\sim 4N_{a}\left| g_{c}\right| ^{2}/\left( \kappa
\gamma _{s}\right) \sim 3\rho _{n}L_{a}/k_{s}^{2}\sim d_{o}$, where $d_{o}$
denotes the on-resonance optical depth of the atomic ensemble
which can be quite large with the current experimental
technology [6-8]. So we can have a considerably large
signal-to-noise ratio for the collective atomic mode even in the
free-space case. A detailed treatment of the free-space
light-atom interaction can be found in Refs. [9,10].

\begin{figure}[tbp]
\epsfig{file=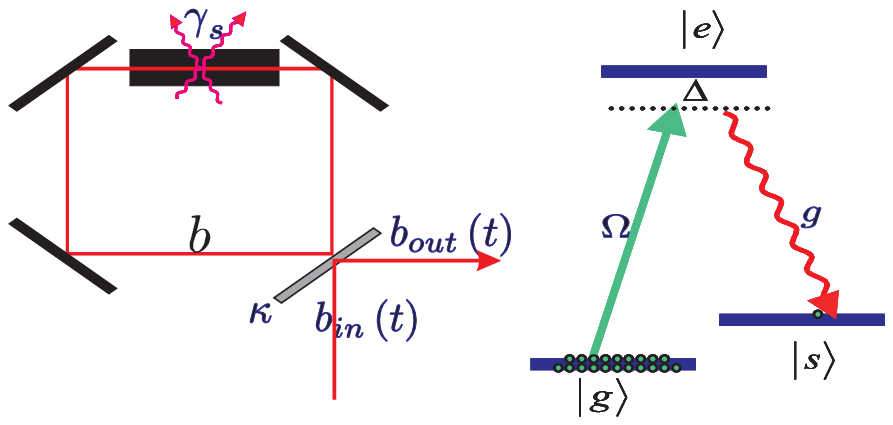,width=8cm} 
\caption{Schematic system configuration}
\end{figure}